\documentclass[twocolumn,amsmath,amssymb,aps,prb]{revtex4}
\usepackage{kotex}
\usepackage{graphicx}
\usepackage{dcolumn}
\usepackage{bm}

\usepackage{physics}
\usepackage{color}

\usepackage{multirow}

\begin{document}


\title{Hund's physics and the magnetic ground state of CrOX (X = Cl, Br)}

\author{Seung Woo Jang$^{\dag}$}
\affiliation{Department of Physics, Korea Advanced Institute of Science and Technology (KAIST), Daejeon 34141, Korea}

\author{Do Hoon Kiem$^{\dag}$}
\affiliation{Department of Physics, Korea Advanced Institute of Science and Technology (KAIST), Daejeon 34141, Korea}

\author{Juhyeok Lee$^{\dag}$}
\affiliation{Department of Physics, Korea Advanced Institute of Science and Technology (KAIST), Daejeon 34141, Korea}

\author{Yoon-Gu Kang}
\affiliation{Department of Physics, Korea Advanced Institute of Science and Technology (KAIST), Daejeon 34141, Korea}

\author{Hongkee Yoon}
\affiliation{Department of Physics, Korea Advanced Institute of Science and Technology (KAIST), Daejeon 34141, Korea}

\author{Myung Joon Han}\email{mj.han@kaist.ac.kr}
\affiliation{Department of Physics, Korea Advanced Institute of Science and Technology (KAIST), Daejeon 34141, Korea}

\begin{abstract}
To understand the magnetic property of layered van der Waals materials CrOX (X = Cl, Br), we performed the detailed first-principles calculations for both bulk and monolayer. We found that the charge-only density functional theory combined with the explicit on-site interaction terms (so-called cDFT$+U$) well reproduces the experimental magnetic ground state of bulk CrOX, which is not the case for the use of spin-dependent density functional (so-called sDFT$+U$). Unlike some of the previous studies, our results show that CrOX monolayers are antiferromagnetic as in the bulk. It is also consistent with our magnetic force linear response calculation of exchange couplings $J_{\rm ex}$. The result of orbital-decomposed $J_{\rm ex}$ calculations shows that the Cr $t_\textrm{2g}$-$t_\textrm{2g}$ component mainly contributes to the antiferromagnetic order in both bulk and monolayer. Our result and analysis show that taking the correct Hund's physics into account is of key importance to construct the magnetic phase diagram and to describe the electronic structure.

\end{abstract}

\maketitle


\section{Introduction}
Magnetism in 2-dimensional (2D) van der Waals (vdW) materials has attracted intensive research attention from the point of view of both fundamental physics and applications \cite{burch_magnetism_2018,gibertini2019magnetic,huang2017E,wei2020emerging,park2016opportunities}. This intriguing family of materials can provide a new platform to explore various magnetism and to realize the next generation storage device \cite{kryder1992magnetic,wolf2001spintronics,li2019intrinsic}. There have been many materials suggested for high critical temperature ($T_\textnormal{C}$) ferromagnetism and other useful spin patterns \cite{huang2017E,bonilla_strong_2018,gong2017E,freitas2015ferromagnetism, jiang2018controlling, deng2018gate, sun2020room, song2018giant, huang2018electrical, klein2018probing,jiang2018electric}. Importantly, however, the characterization of 2D magnetism is more challenging compared to their bulk counterparts. For example, the identification of the ground state spin order cannot rely, in many situations, on the standard techniques such as neutron scattering due to the small sample size. And, in this regard, the role of theoretical calculation becomes even more important. In particular, the first-principles calculations have been playing a critical role not only in predicting and designing new 2D magnetic materials but also in characterizing the magnetic properties.

Density functional theory (DFT) provides the overarching theoretical framework in this type of approach. In DFT calculations of 2D vdW materials, the choice of `exchange-correlation (XC) functional' requires more care. It is partly because the XC functional for vdW interaction has not yet been quite well established \cite{berland2015van}. It is also related to the notorious problem in describing `strongly correlated electron materials' within first-principles scheme \cite{martin_int_book2016,anisimov_cor_book2010, kotliar2006electronic, anisimov1997first}. The standard approximations such as LDA (local density approximation) and GGA (generalized gradient approximation) are known to have severely limited capability when the localized $d$ or $f$ orbitals are partially filled, which is indeed often the case for magnetic materials. Further, the theoretical simulation can become even more difficult as the information from experiments (e.g., lattice parameter and the size of moment, etc) is typically less than the bulk situation. Thus, we believe, the rise of 2D vdW magnetic material can pose a challenge to the first-principles methodology. A more detailed understanding, investigation, and hopefully the meaningful revision of the widely-used methods are strongly requested.

In the present study, we investigate CrOX (X = Br and Cl) for which the previous theoretical predictions are controversial. A series of DFT$+U$ calculations predicted that the monolayer CrOX is ferromagnetic (FM) with $T_\textnormal{C}$ as high as 160 K \cite{CrOX2018T,CrCX2019T,qing2020magnetism,nair2020bi}. On the other hand, a recent hybrid functional calculation of HSE06 found the antiferromagnetic (AFM) ground state \cite{zhang_super-exchange_2019}. This situation, on one hand, demonstrates the difficulty in predicting the magnetic property of 2D vdW materials even within the standard methodological frame. On the other hand, it also shows that further understanding and the detailed investigation of the working principles of these approximations are required.

In the below, we try to provide useful insights on these issues. We show that the FM ground state obtained by previous DFT$+U$ is essentially attributed to the unphysical spin density contribution which is not properly described and largely double-counted within their $+U$ formalisms. The use of charge-only density for the `uncorrelated' Hamiltonian can remedy this problem as observed recently in other magnetic materials \cite{chen_spin-density_2016,park_density_2015,ryee_effect_2018,jang_charge_2018,jang_microscopic_2019}. Our calculation based on charge-only density functional theory plus $U$ (cDFT$+U$) shows that the ground state spin order is AFM for both CrOBr and CrOCl in good agreement with the HSE06 result \cite{zhang_super-exchange_2019}. The electronic structure analysis clearly shows that the correct Hund's physics is captured by cDFT$+U$. We also provide the detailed information of magnetic couplings; while some orbital contributions mediating the FM couplings are certainly enhanced when the system becomes 2D (i.e., in monolayer), the ground state configuration is AFM being consistent with the total energy results.

\begin{figure*}[t]
	\centering
	\includegraphics[width=1\linewidth]{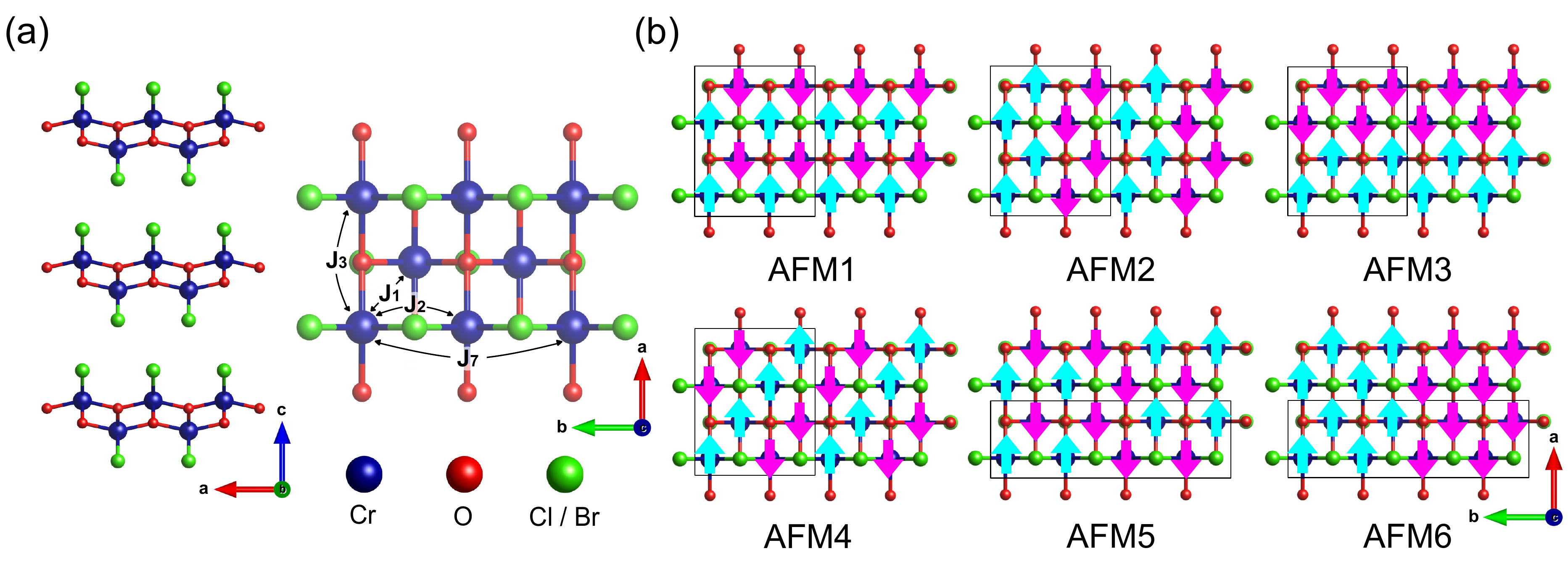}
	\caption{(a) The top and side view of CrOX crystal structure. $J_n$ refers to the $n$-th neighbor coupling constant. (b) Six different AFM spin order that can be realized within the 2$\times$2$\times$1 and 1$\times$4$\times$1 supercell (see the black-solid rectangles). The up (light blue) and down (magenta) arrows represent the spin directions.}
	\label{fig:figure1}
\end{figure*}

\section{Computational DETAILS}

Density functional theory plus $U$ (DFT+$U$) calculations were carried out using Vienna \textit{ab initio} simulation package (VASP) based on projector augmented wave (PAW) potential \cite{kresse_efficient_1996, kresse_efficiency_1996} and within Perdew-Burke-Ernzerhof (PBE) type of GGA  functional \cite{perdew_generalized_1996}. The `DFT+D3' type of vdW correction was adopted for bulk calculations to properly describe the inter-layer interactions \cite{grimme_consistent_2010,grimme_effect_2011}. We will strictly distinguish spin-polarized GGA+$U$ (SGGA$+U$) from spin-unpolarized GGA+$U$ (or cDFT$+U$) following Ref.~\onlinecite{ryee_effect_2018,park_density_2015,chen_spin-density_2016,jang_charge_2018}. So-called fully localized limit of (S)GGA$+U$ functional was adopted as suggested by Liechtenstein and co-workers \cite{liechtenstein_density-functional_1995,czyzyk_local-density_1994,anisimov_density-functional_1993,solovyev_corrected_1994}.
The $\Gamma$-centered k-grid for monolayer and bulk was 8$\times$26$\times$1 and 8$\times$26$\times$7, respectively, for 1$\times$4$\times$1 supercell. For the 2$\times$2$\times$1 supercell calculations, we used 13$\times$16$\times$1 and 13$\times$16$\times$7, respectively. The optimized crystal structures (monoclinic for CrOCl and orthorhombic for CrOBr) were used with the force criteria of 5$\times10^{-3}$ eV/\AA. We find that the use of experimental structures does not change any of our conclusions. For monolayers, a  20 \AA~ vacuum space was taken into account. Plane-wave energy cutoff is 550 eV. Constrained random phase approximation (cRPA) calculations \cite{aryasetiawan_frequency-dependent_2004} were performed to estimate Hubbard $U$ and Hund $J_{\rm Hund}$ interaction parameters by using `ecalj' software package \cite{ecalj}. So-called `$d$ model' downfolding was adopted with maximally localized Wannier function (MLWF) technique \cite{sasioglu_effective_2011,sakuma_first-principles_2013,jang_direct_2016}. 
Magnetic force linear response calculations \cite{liechtenstein_local_1987,han_electronic_2004, yoon_reliability_2018,yoon2020jx} are performed with MLWF projections \cite{mostofi_updated_2014,pizzi_wannier90_2019} to estimate the inter-atomic magnetic exchange interactions, $J_{\rm ex}$, which is defined as
\begin{equation} \label{eq:1}
H=-  \sum_{i,j} \hat{e}_{i} J_{\rm ex}  \hat{e}_{j} 
\end{equation}
where $\hat{e}_{i,j}$ is the unit spin vector at the atomic site $i$ and $j$.

\section{RESULT and Discussion}

CrOX is a layered vdW material with orthorhombic structure ($Pmmn$, space group no. 59) \cite{angelkort_observation_2009, CrOCl_1975}. As shown in Fig.~\ref{fig:figure1}(a), Cr atoms are located in the octahedral anion cage which is composed of four oxygens and two X. CrOX bulk is known to be AFM insulators \cite{angelkort_observation_2009, CrOCl_1975, coic_chromium_1981,zhang2019magnetism}. Interestingly, previous studies report the two different AFM spin order for bulk CrOCl, namely AFM2 and AFM5 (see Fig.~\ref{fig:figure1}(a)) \cite{angelkort_observation_2009, CrOCl_1975}. According to the more recent experiment \cite{angelkort_observation_2009}, the AFM2 order is accompanied by monoclinic distortion ($\alpha=90.06^\circ$). The layered structure and the weak inter-layer interaction implies the relatively easy exfoliation, which triggers recent investigations of 2D vdW magnetism in this material.

\begin{figure*}
	\centering
	\includegraphics[width=1\linewidth]{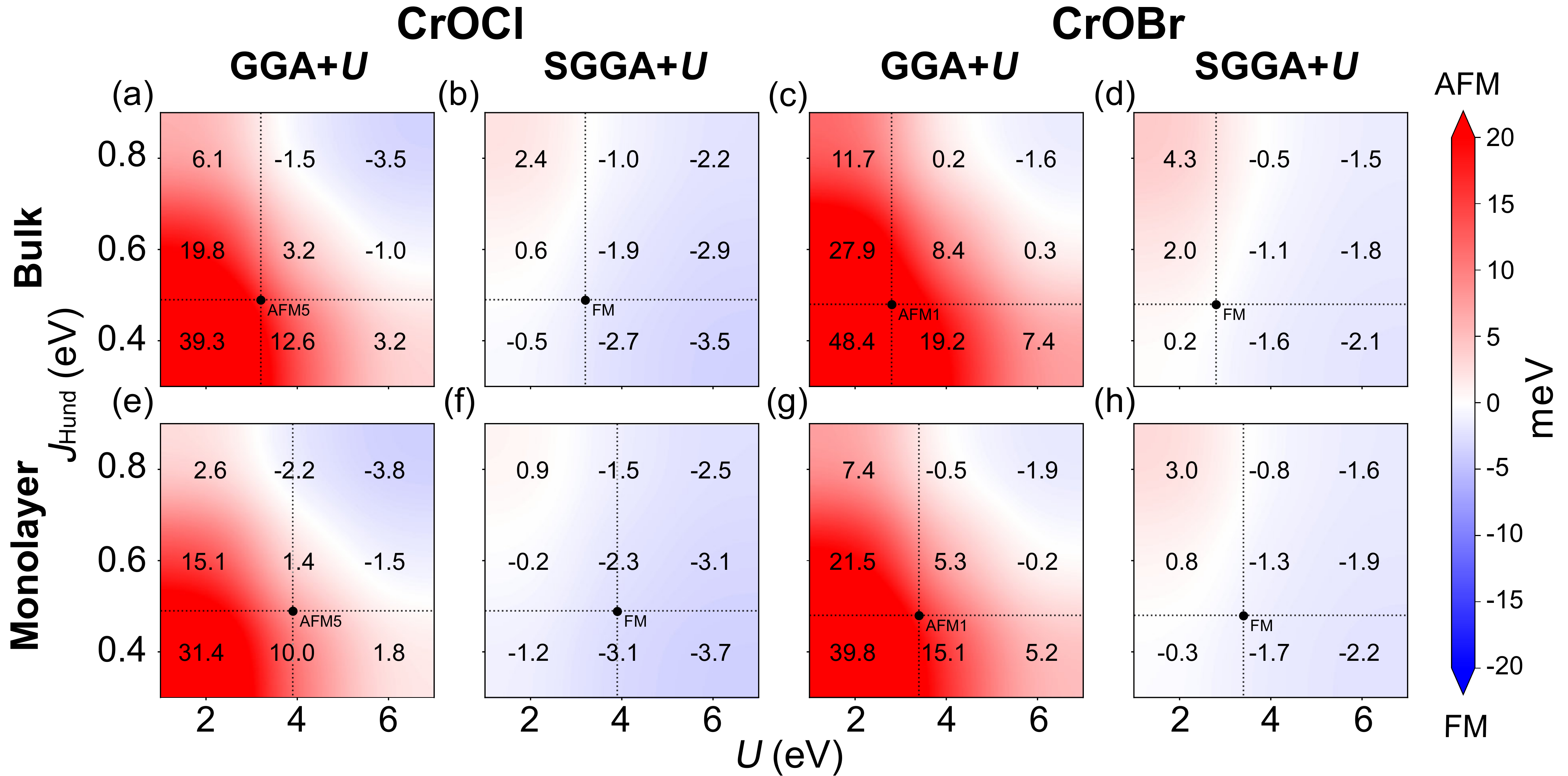}
	\caption{The magnetic phase diagrams for (a, b, e, f) CrOCl and (c, d, g, h) CrOBr. The calculated total energy differences between FM and AFM, E$_{\rm FM} -$E$_{\rm AFM}$, are presented as a function of Hubbard $U$ (horizontal axis) and Hund $J_{\rm Hund}$ (vertical axis). The results from GGA+$U$ and SGGA$+U$ are given in (a, c, e, g) and in (b, d, f, h), respectively. Bulk and monolayer results are presented in (a-d) and (e-h), respectively. The numbers inside the phase diagrams show the calculated energy differences in the unit of meV/Cr. The blue and red colors denote the FM and AFM spin ground state, respectively (see the color bar).}
	\label{fig:figure2}
\end{figure*}

Notably, a series of previous theoretical studies predicted the 2D monolayer ferromagnetism for both CrOBr and CrOCl \cite{Mazari2018T,CrOX2018T,CrCX2019T,nair2020bi,qing2020magnetism}. The calculated total energy of FM order is predicted to be lower than that of AFM by about 2--15 meV per Cr \cite{Mazari2018T,CrOX2018T,CrCX2019T,nair2020bi,qing2020magnetism}. It is particularly noted that GGA ($U$ = 0)\cite{Mazari2018T}, Dudarev's form of SGGA$+U$ ($U_{\rm eff}$ = 3, 5, 7 eV in Ref.\onlinecite{CrOX2018T}; $U_{\rm eff}$=  7.0 eV in Ref.\onlinecite{nair2020bi}) 
and Liechtenstein functional ($U$=3.89 eV and $J_{\rm Hund}$ = 0.98 eV; performed only for CrOCl ~\cite{CrCX2019T})  coincidently give the same conclusion of FM order.
On the contrary, a more recent calculation of using HSE06 functional reported the AFM spin ground state being more stable than FM by $\sim$2 meV/Cr ~\cite{zhang_super-exchange_2019}.

Given that many of conventional experimental tools cannot be utilized for probing 2D magnetism because of the small sample size and thus the characterization of magnetic order largely relies on theoretical calculation, establishing the reliability and the predictive power of the widely-used computation methods is of critical importance. Also from the theoretical point of view, urgently required is to have a detailed understanding of the working principles of widely-used standard functionals.

\subsection{SGGA$+U$}

\begin{table}[b]
	\caption{
		The cRPA calculation results of $U$ and $J_{\rm Hund}$ for CrOX bulk and monolayer. The unit is eV.
	}
	\begin{ruledtabular}
		\begin{tabular}{lldd}
			\multicolumn{2}{c}{}&\multicolumn{1}{c}{$U$}&\multicolumn{1}{c}{$J_{\rm Hund}$}\\ 
			\hline
			\multirow{2}{*}{CrOCl} &bulk& 3.2 & 0.49\\
			&monolayer & 3.9 & 0.49 \\ \hline
			\multirow{2}{*}{CrOBr} &bulk& 2.8 & 0.48\\
			&monolayer & 3.5 & 0.48 \\
		\end{tabular}
	\end{ruledtabular}
	\label{tab:table1}
\end{table}

\begin{figure*}[t]
	\centering
	\includegraphics[width=0.8\linewidth]{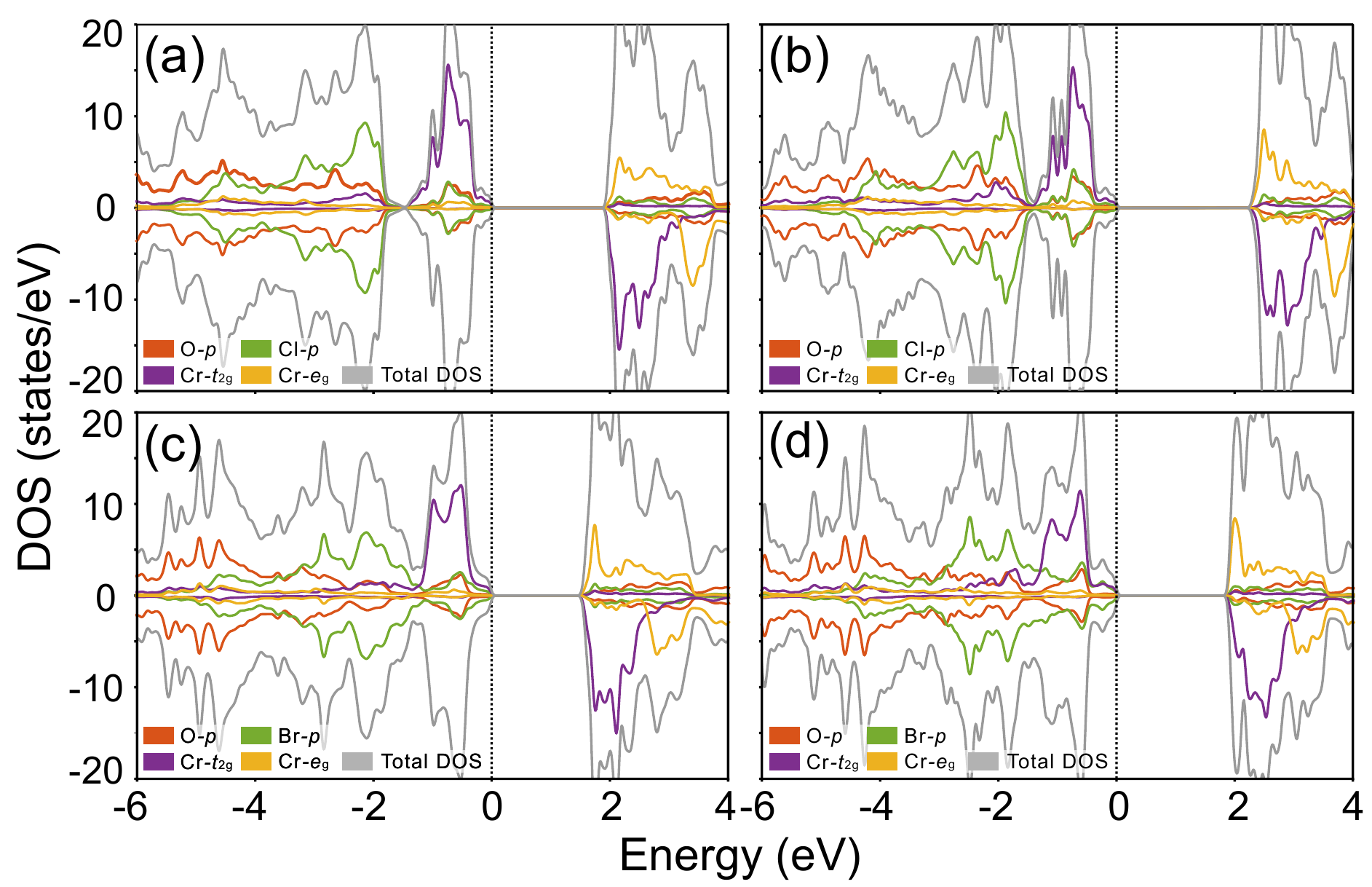}
	\caption{The calculated DOS of (a)/(c) bulk and (b)/(d) monolayer CrOCl/CrOBr where cRPA values of $U$ and $J_\textrm{Hund}$ were used. The most stable spin configurations are presented as obtained using GGA+$U$; namely, AFM5 and AFM1 for CrOCl and CrOBr, respectively. The red, green, violet, and yellow lines denote the O-$p$, X-$p$, Cr-$t_\textrm{2g}$, and Cr-$e_\textrm{g}$ orbital state, respectively. The positive and negative panels denote the up and down spin DOS, respectively. Fermi energy is set to zero.}
	\label{fig:figure3}
\end{figure*}

As the first step, we calculated the total energy differences, $E_{\rm FM}-E_{\rm AFM}$, within SGGA$+U$ and in a wide range of $U$ and $J_{\rm Hund}$. The results are presented in Fig.~\ref{fig:figure2}. Since Dudarev functional corresponds to the $J_{\rm Hund}$ = 0 limit of Liechtenstein functional \cite{ryee_effect_2018}, we adopted the more general form of Liechtenstein functional and varied both $U$ and $J_{\rm Hund}$. Our results for bulk/monolayer CrOCl and CrOBr are summarized in Fig.~2(b)/(f) and (d)/(h), respectively. The monolayer CrOX is predicted to be FM in a wide range of parameter space; see Fig.~2(f) and (h). In particular, the small $J_{\rm Hund}$ region is FM regardless of $U$ values. Our result is therefore consistent with the previous calculation by Miao {\it et. al.}, which corresponds to $J_{\rm Hund}$ = 0 eV points of our phase diagram \cite{CrOX2018T}, and that by Mounet {\it et. al.} \cite{Mazari2018T} where $U=J_{\rm Hund}$ = 0. At $U$ = 3.98 eV and $J_{\rm Hund}$ = 0.98 eV (not shown), our calculation also gives the FM ground state as reported by  Wang {\it et. al.} \cite{CrCX2019T}.

The problem is, however, that SGGA and SGGA$+U$ with these parameters predict the FM ground state also for the bulk CrOX which is in sharp contrast to the experimental fact \cite{angelkort_observation_2009, CrOCl_1975, coic_chromium_1981}. The calculated bulk phase diagram is shown in Fig.~\ref{fig:figure2}(b) and (d) for CrOCl and CrOBr, respectively, which are quite similar to those of monolayer. In order to stabilize the correct bulk AFM spin order, $J_{\rm Hund}$ should be strong ($\geq$ 0.6 eV) and simultaneously Hubbard $U$ be small enough ($\leq$ 2 eV). On top of it, in order to have monolayer ferromagnetism, $J_{\rm Hund}$ should get reduced and $U$ enlarged as the system dimension goes to the 2D limit, which seems quite unlikely.

In order to estimate the realistic interaction strengths, we performed the cRPA calculation \cite{aryasetiawan_frequency-dependent_2004,sasioglu_effective_2011}, and the results are summarized in Table~\ref{tab:table1} (also indicated by dashed lines in Fig.~2). As somewhat expected, the Hubbard $U$ is enhanced when the system dimension is reduced from bulk to monolayer while $J_{\rm Hund}$ remains the same. This cRPA result indicates that SGGA$+U$ fails to reproduce the AFM ground state for bulk CrOX.

\subsection{GGA$+U$}

In fact, the limitation of SGGA$+U$ has recently been pointed out in literature\cite{ryee_effect_2018, chen_spin-density_2016, park_density_2015,jang_charge_2018}. It is attributed to the unphysical description of magnetic exchange; when SGGA or LSDA is combined with $+U$ functionals, the Hund's physics is not properly controlled and likely gives rise to the unrealistic solution \cite{ryee_effect_2018, chen_spin-density_2016, park_density_2015,jang_charge_2018}. Keeping this point in our mind, we performed (spin unpolarized) GGA$+U$ calculations whose results are presented in Fig.~\ref{fig:figure2}(a)/(e) and (c)/(g) for bulk/monolayer CrOCl and CrOBr, respectively. Note that this functional can still describe the magnetic solutions while the interaction parts, namely the $U$--$J_{\rm Hund}$ functionals, are solely responsible for magnetism as can be seen explicitly in its interaction functional:
\begin{eqnarray}  
	E^{\textrm{int}}_{\textrm{FLL}}  &=& \frac{1}{2}\sum_{\{m_i\},\sigma,\sigma'} \{n^{\sigma\sigma}_{m_1m_2}\langle m_1,m_3|V_{ee}|m_2,m_4 \rangle n^{\sigma'\sigma'}_{m_3m_4}\\
	 &-& n^{\sigma\sigma'}_{m_1m_2} \langle m_1,m_3|V_{ee}|m_4,m_2 \rangle n^{\sigma'\sigma}_{m_3m_4} \},
\end{eqnarray}
where $n^{\sigma\sigma'}_{m_1m_2}$ are the elements of on-site density matrix $\mathbf{n}$ for orbitals $\{m_i\}$ and spins $\sigma,\sigma'$.
The matrix elements of on-site Coulomb interaction are expressed by
\begin{eqnarray} 
& &	\langle m_1,m_3|V_{ee}|m_2,m_4 \rangle = \sum_{\{m_i'\}}\Big[S_{m_1m_1'}S_{m_3m_3'}\\
	&\times& \Big\{\sum_{k=0}\alpha_k(m_1',m_3',m_2',m_4')F^k\Big\} S^{-1}_{m_2'm_2}S^{-1}_{m_4'm_4} \Big]
\end{eqnarray}
where $\alpha_k$ and $F^k$ refers to Racah-Wigner numbers and Slater integrals, respectively, and $S$ is a transformation matrix from spherical harmonics to the predefined local basis sets. 
From the conventional expression, $U=F^0$, $J=(F^2+F^4)/14$, and $F^4/F^2=0.625$ for $d$-orbitals. See, for more details, Ref.~ \onlinecite{ryee_effect_2018} which includes discussion regarding both FLL (fully localized limit) and AMF (around mean-field) formalism. 
Fig.~\ref{fig:figure2}(b, f) and (d, h) clearly show that the ground state profile as a function of $U$ and $J_{\rm Hund}$ is qualitatively different from that of SGGA$+U$. In GGA$+U$ calculations, the FM region is located in the large $J_{\rm Hund}$ and large $U$ region for both CrOCl and CrOBr while the small $J_{\rm Hund}$ and small $U$ enhance the stability of AFM solutions relative to FM.

Importantly, GGA$+U$ gives rise to the correct AFM ground state for bulk CrOCl and CrOBr. The points corresponding to cRPA parameters of $U$=3.2 (2.8) eV and $J_{\rm Hund}$=0.49 (0.48) eV for CrOCl (CrOBr) are located deep inside the AFM phase as shown in Fig.~\ref{fig:figure2}(a) (Fig.~2(c)). 
The calculated total energies are summarized in Table~\ref{tab:table_totalE}.
Contrary to SGGA$+U$, GGA$+U$ predicts the AFM ground state also for monolayers. In the below, we argue that GGA$+U$ is more reliable not simply because its result for bulk CrOX correctly reproduces the AFM ground state, but also because it describes the correct Hund physics for this multi-band magnetic material.

\begin{table}[ht]
\caption{The calculated relative total energy by GGA+$U$ with the cRPA parameters of $U$ and $J_{\rm Hund}$. The unit is meV/Cr.}
\begin{tabular}{llccccccc}
\hline
\hline
                       &      & FM   & AFM1 & AFM2 & AFM3 & AFM4 & AFM5 & AFM6 \\ \hline
\multirow{2}{*}{CrOCl} & bulk & 13.8 & 2.6  & 0.3  & 12.3 & 4.3  & 0.0  & 5.2  \\ \cline{2-9} 
                       & mono & 6.4  & 1.6  & 0.3  & 10.2 & 6.1  & 0.0  & 2.2  \\ \hline
\multirow{2}{*}{CrOBr} & bulk & 21.7 & 0.0  & 4.5  & 12.2 & 8.3  & 0.8  & 11.3 \\ \cline{2-9} 
                       & mono & 12.3 & 0.0  & 3.3  & 11.1 & 9.0  & 0.7  & 6.6  \\ \hline
\hline
\end{tabular}
\label{tab:table_totalE}
\end{table}

\begin{figure}[t]
	\centering
	\includegraphics[width=1\linewidth]{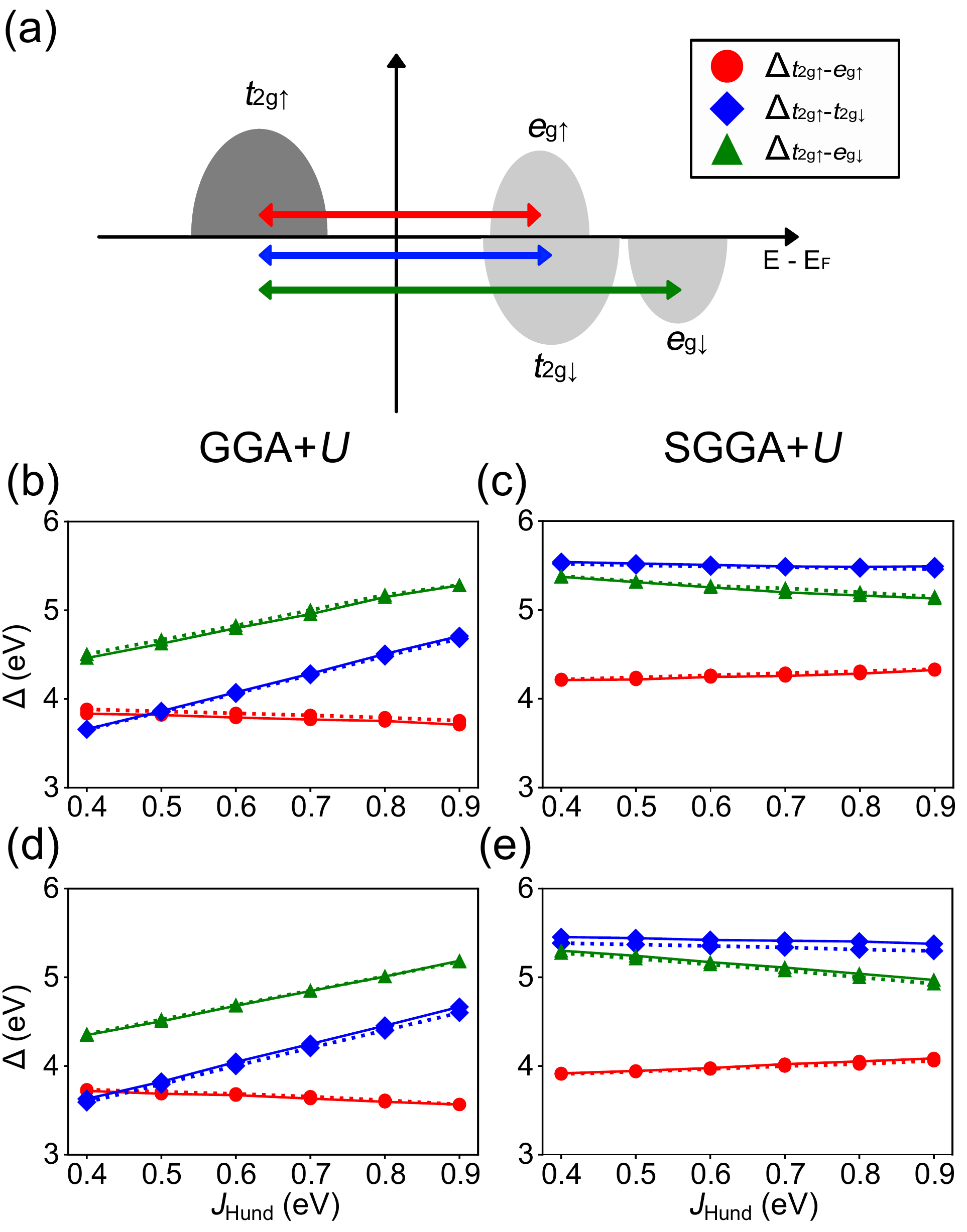}
	\caption{(a) A schematic diagram for Cr-$e_{g}$ and $t_{2g}$ PDOS. (b)--(e) The calculated charge excitation energy $\Delta$ as a function of $J_{\rm Hund}$; (b)/(d) GGA+$U$ and (c)/(e) SGGA+$U$ for CrOCl/CrOBr. The values from cRPA were used. The dashed and solid lines denote the results of bulk and monolayer, respectively.}
	\label{fig:figure4}
\end{figure}

\subsection{Electronic structure analysis}

Fig.~3(a, b) and (c, d) shows the projected density of states (PDOS) for CrOCl and CrOBr, respectively. The violet and yellow lines represent Cr-$t_{2g}$ and $e_g$ states, respectively. The main difference between CrOCl and CrOBr is the reduced band gap in the case of X=Cl by $\sim$0.3 eV. Here it should be noted that different $U$ and $J_{\rm Hund}$ values were used for each material as obtained from cRPA calculation (see Table~I). The electronic structure of the monolayer is similar to that of bulk while the up-spin $t_{2g}$ bands are closer to the lower-lying anion-$p$ states in monolayer. Other than these rather minor differences, the PDOS of these four different systems are essentially the same; they all share the same $d^3$ configuration of Cr$^{3+}$ carrying $t_{2g}^{\uparrow,3} ~ t_{2g}^{\downarrow,0} ~  e_{g}^{\uparrow\downarrow,0}$ (where the arrows represent the spin directions). This electronic structure is schematically represented in Fig.~4(a) where we denote the energy separation between two given states by $\Delta$;
$\Delta_{\textit{t}_{2g}^\uparrow-\textit{e}_g^\uparrow}$,
$\Delta_{\textit{t}_{2g}^\uparrow-\textit{t}_{2g}^\downarrow}$, and
$\Delta_{\textit{t}_{2g}^\uparrow-\textit{e}_g^\downarrow}$ refers to
the electron excitation energy from $\textit{t}_{2g}^{\uparrow}$  to 
$\textit{e}_{g}^{\uparrow}$, $\textit{t}_{2g}^{\downarrow}$, and $\textit{e}_{g}^{\downarrow}$, respectively.

Now we estimated $\Delta$'s based on the PDOS results of GGA$+U$ and SGGA$+U$. Namely, $\Delta$ is given by
	\begin{equation}
	\Delta=\frac{\int_{E_f}^{E_f+y} Eg_{\alpha}(E)dE}{\int_{E_f}^{E_f+y}g_{\alpha}(E)dE}-\frac{\int_{E_f-x}^{E_f}Eg_{\alpha}(E)dE}{\int_{E_f-x}^{E_f}g_{\alpha}(E)dE}
	\end{equation} 
where $\alpha$ is the spin-orbital index and $E_f$ the Fermi energy. $g_{\alpha}(E)$ denotes the calculated PDOS, and $x$ and $y$ need to be properly chosen to include the desired Cr states. We used $x=$2.0 and $y$=5.0 eV for all cases. Fig.~\ref{fig:figure3} shows that these values reasonably well cover the main Cr PDOS.

The calculated $\Delta$'s as a function of $J_{\rm Hund}$ are summarized in Fig.~4(b)-(e). Within GGA$+U$,
$\Delta_{\textit{t}_{2g}^\uparrow-\textit{t}_{2g}^\downarrow}$ (blue diamonds) and
$\Delta_{\textit{t}_{2g}^\uparrow-\textit{e}_g^\downarrow}$ (green triangles)
increase as $J_\textrm{H}$ increases while the change of
$\Delta_{\textit{t}_{2g}^\uparrow-\textit{e}_g^\uparrow}$ (red circles) is negligible; see Fig.~3(b) and (d). On the other hand, the opposite trend is clearly observed in SGGA$+U$. Namely,
$\Delta_{\textit{t}_{2g}^\uparrow-\textit{t}_{2g}^\downarrow}$ and
$\Delta_{\textit{t}_{2g}^\uparrow-\textit{e}_g^\downarrow}$
are slightly but gradually reduced as a function of $J_\textrm{Hund}$.
Hund’s rule tells us that this energy splitting between the opposite spins should be enlarged as $J_{\rm Hund}$ increases.
Note that only GGA$+U$ reproduces this correct Hund's physics; {\it i.e.}, the equal spin occupation is energetically favored and its energy scale is determined by $J_{\rm Hund}$. It is therefore the key required feature from the methodological point of view. This physically correct Hund's physics is captured by GGA$+U$ but not by SGGA$+U$.

This analysis of $J_{\rm Hund}$ dependence provides further understanding of the different phase diagrams by GGA$+U$ and SGGA$+U$ (shown in Fig.~\ref{fig:figure2}). Since the superexchange interaction is expressed by $t^2 / \Delta$, the magnetic exchange coupling $J_{\rm ex}$ is inversely proportional to $\Delta$; $J_{\rm ex} \sim \Delta^{-1}$. As a result, the increasing trend of 
$\Delta_{\textit{t}_{2g}^\uparrow-\textit{t}_{2g}^\downarrow}$  and
$\Delta_{\textit{t}_{2g}^\uparrow-\textit{e}_g^\downarrow}$ as a function of $J_{\rm Hund}$ naturally gives the weaker AFM coupling in the larger $J_{\rm Hund}$ region. It renders the FM order stabilized in the large $J_{\rm Hund}$ regime of GGA$+U$ phase diagram.

From the point of view of Hund's rule, the electronic behavior described by SGGA$+U$ is unphysical, and it is originated from the magnetic contributions intrinsically present in SGGA as discussed in recent literature \cite{ryee_effect_2018, chen_spin-density_2016, park_density_2015}. Note that, in the limit of $U\rightarrow 0$ and $J_{\rm Hund}\rightarrow 0$, the ground state solution of SGGA$+U$ is FM; see Fig.~\ref{fig:figure2}. It demonstrates the presence of intrinsic magnetic energy scale in SGGA$+U$ (often denoted by Stoner $I$) which competes with Hund energy scale $J_{\rm Hund}$ \cite{ryee_effect_2018, jang_charge_2018, chen_spin-density_2016, park_density_2015}.

\begin{table}[t]
	\caption{The calculated magnetic coupling constants $J_{\rm ex}$ and their orbital decompositions as estimated by MFT. The four major interactions $J_{1}$, $J_{2}$, $J_{3}$, and $J_{7}$ are defined in Fig.1(a). Here $t_{\rm 2g}$-$e_{\rm g}$ interactions refer to the summation of $t_{\rm 2g}$-$e_{\rm g}$ and $e_{\rm g}$-$t_{\rm 2g}$ contributions. The unit is meV.}
	\begin{ruledtabular}
		\begin{tabular}{llldddd}
			\multicolumn{2}{c}{}&\multicolumn{1}{c}{interaction}&\multicolumn{1}{c}{$J_1$}&\multicolumn{1}{c}{$J_2$}&\multicolumn{1}{c}{$J_3$}&\multicolumn{1}{c}{$J_7$}\\ 
			\hline
			CrOCl &bulk	          &total     					   & -0.9 & -1.1&0.4&-0.2\\\cline{3-7}
			&                 &$t_\textrm{2g}$-$t_\textrm{2g}$     & -1.7 & -1.7&-1.3&-0.2\\
			&                 &$t_\textrm{2g}$-$e_\textrm{g}$       & 0.8 & 0.6 &1.8&0.0\\
			&                 &$e_\textrm{g}$-$e_\textrm{g}$        & 0.0 & 0.0 &-0.1&0.0\\\cline{2-7}
			&monolayer &total     							       & -0.4 & -0.5&0.8&-0.2\\\cline{3-7}
			&                 &$t_\textrm{2g}$-$t_\textrm{2g}$     & -1.2 & -1.1&-0.9&-0.2\\
			&                 &$t_\textrm{2g}$-$e_\textrm{g}$       & 0.8 & 0.6&1.8&0.0\\
			&                 &$e_\textrm{g}$-$e_\textrm{g}$        &  0.0 & 0.0&-0.1&0.0\\\hline
			CrOBr &bulk	         &total     					  & -1.9 & -0.5& 0.3&-0.2\\\cline{3-7}
			&                 &$t_\textrm{2g}$-$t_\textrm{2g}$     & -2.5 & -0.9&-1.6 &-0.2\\
			&                 &$t_\textrm{2g}$-$e_\textrm{g}$       & 0.6  & 0.4 & 2.0 &0.0\\
			&                 &$e_\textrm{g}$-$e_\textrm{g}$        & 0.0  & 0.0  & -0.1 &0.0\\\cline{2-7}
			&monolayer &total     							       & -1.2 & -0.2 &0.6&-0.2\\\cline{3-7}
			&                 &$t_\textrm{2g}$-$t_\textrm{2g}$     & -1.8 & -0.6 &-1.1&-0.2\\
			&                 &$t_\textrm{2g}$-$e_\textrm{g}$       & 0.6 & 0.4 &1.8  &0.0\\
			&                 &$e_\textrm{g}$-$e_\textrm{g}$        & 0.0 & 0.0 &-0.1 &0.0\\
		\end{tabular}
	\end{ruledtabular}
	\label{tab:table3}
\end{table}

\subsection{Magnetic coupling constants}

In order to estimate the magnetic interactions, we performed the magnetic force theory (MFT) calculations \cite{liechtenstein_local_1987,han_electronic_2004, yoon_reliability_2018} whose
results are summarized in Table~\ref{tab:table3}.
For bulk, the inter-layer $J_{\rm ex}$ is negligibly small $\sim$0.01 meV (not shown). We also found that  $J_{n=4, 5, 6}$ and $J_{n\geq 8}$ are less than $\sim$0.05 meV. 
Among the three major interactions, $J_{1}$ and $J_{2}$ are AFM, and $J_{3}$ is FM. The former and the latter is reduced and enhanced, respectively, in monolayer. Out of the competitions of these couplings, the ground state spin configuration is largely determined.
Interestingly, however, the noticeable coupling ($\sim$0.2 meV) is also found in the seventh-neighbor interactions, $J_{7}$, which plays a role in stabilizing the `double-stripe' spin patterns (see Fig.~\ref{fig:figure1}.)

Among all possible spin configurations within 4$\times$8$\times$1 magnetic supercell, the most stable phase is AFM5 and AFM1 for X=Cl (bulk and monolayer) and Br (bulk and monolayer), respectively, in good agreement with the total energy calculation (see Fig.~\ref{fig:figure2}). The main difference between CrOCl and CrOBr is the relative strength between $J_1$ and $J_2$. In CrOCl, the AFM5 and AFM2 order is the first and the second most stable phase, respectively. Since the energy difference $E_{\rm AFM5}-E_{\rm AFM2}$ is proportional to $J_{1}-J_{2}+2J_{7}$ and $(J_{1}-J_{2})\sim$0.1 meV (in both bulk and monolayer), $J_{7}$ plays an important role. In CrOBr, on the other hand, AFM1 and AFM5 are the first and the second most stable phase, respectively. Given that $E_{\rm AFM5}-E_{\rm AFM1}$ is proportional to $-J_{1}+J_{2}+2J_{7}$ and that $(-J_{1}+J_{2})\sim$1.0 meV, the effect of $J_{7}$ is not significant. 

Further insight can be obtained from the orbital decomposition of $J_{\rm ex}$. Table~\ref{tab:table3} shows that $J_{\rm ex}(t_{\rm 2g}-t_{\rm 2g})$ is responsible for the inter-atomic AFM couplings while $J_{\rm ex}(t_{\rm 2g}-e_{\rm g})$ is mostly FM. The interactions between $e_{\rm g}$-$e_{\rm g}$ orbitals are much smaller and therefore can hardly contribute to the spin order as expected from the electronic configuration. Therefore the magnetic ground state of CrOX can be understood as the result of the competition between $J_{\rm ex}(t_{\rm 2g}$-$t_{\rm 2g})$ and $J_{\rm ex}(t_{\rm 2g}$-$e_{\rm g})$. 


\section{Summary}
We investigated the magnetic property of bulk and monolayer CrOX for which previous theoretical calculations provide the controversial predictions. We carefully examined the different versions of DFT$+U$ as well as the electronic structure, and the inter-atomic and inter-orbital magnetic couplings. Our results show that the correct Hund's physics is captured only when the spin density is properly taken into account. With cDFT$+U$, the bulk AFM spin order is correctly reproduced. While the FM inter-orbital channels are enhanced in monolayers than in bulk, the ground state spin order is still AFM which is in good agreement with the previous HSE06 result but not with the SDFT$+U$. Our current study provides the useful understanding of the working principles of the widely-used density functional methods as well as the magnetic property of this intriguing vdW materials.

\section{Acknowledgement}
{}$^{\dag}$The first three authors equally contribute to this work.  This work was supported by the National Research Foundation of Korea (NRF) grant funded by the Korea government (MSIT) (No. 2018R1A2B2005204 and NRF-2018M3D1A1058754).
This research was supported by the KAIST Grand Challenge 30 Project (KC30) in 2020 funded by the Ministry of Science and ICT of Korea and KAIST (N11200128).
This work was supported by the National Supercomputing Center with supercomputing resources including technical support (KSC-2020-CRE-0084).




\end{document}